\documentclass[11pt] {article}      
\usepackage{amsmath,amssymb, amsthm, eucal, accents}  
\usepackage{graphicx} 
\usepackage{txfonts}
\usepackage[T1]{fontenc}     
\usepackage{float}                 
\usepackage{mathtools}
\theoremstyle{plain}
\newtheorem{theorem}{Theorem}[section]            
\newtheorem{proposition}[theorem]{Proposition}  
\newtheorem{lemma}[theorem]{Lemma}
\newtheorem{corollary}[theorem]{Corollary}	      %

\theoremstyle{definition}
\newtheorem{definition}[theorem]{Definition}

\numberwithin{theorem}{section}
\numberwithin{equation}{section}
\numberwithin{figure}{section}
\def\q{\partial}
\def\rmid{\hbox{\rm id}}
\usepackage{titlesec}                                                           
\titlelabel{\thetitle.\ }                                                           
\titleformat*{\section}{\fontsize{14pt}{14pt} \bf}                   
\usepackage{fancyhdr}
\def\QED{ $\square$}

\def\Torq{\hbox{\rm Torq}\,}
\def\Curv{\hbox{\rm Curv}\,}

\def\sprl{\ldb}  \def\sprr{\rdb}
\def\TTT{{\hhat T}}
\def\SSS{{\hhat S}}

\def\XX{\mathcal{X}}
\def\FF{\mathcal{F}}
\def\DD{\mathcal{D}}
\def\TT{\mathcal{T}}
\def\SS{\mathcal{S}}
\def\La{\Lambda}
\def\Ker{\hbox{Ker}\,}
\def\caps{\sc}
\def\w{\wedge}
\def\Ld{\hbox{\rm\pounds}^{}}  
\def\zzz{\to}
\def\EE{\mathcal E}
\def\rrho{\varrho}

\def\ra{\rangle}
\def\sprl{[\![}
\def\sprr{]\!]}
\def\ot{\leftarrow}

\newcommand{\jjj}{\,\vrule width6pt height.4pt depth0pt
        \vrule height6pt width.4pt depth0pt\,}    
\def\zza{\mathop{\hbox to 31pt{\rightarrowfill}}}
\def\zzzz{\mathop{\hbox to 49pt{\rightarrowfill}}}
\def\zzzu#1{\smash{\mathop{\zzzz}\limits^{#1}}}
\def\zzu#1{\smash{\mathop{\zza}\limits^{#1}}}
\def\ddd{\Big\downarrow}  
\def\dddr#1{\ddd\rlap{$\vcenter{\hbox{$\scriptstyle#1$}}$}}

\begin{document}

\title{\bf  Relativistic observer and Maxwell's equations:
an example of a non-principal Ehresmann connection}

\author{Jerzy Kocik
\footnote{Preprint P-98-10-029 of the Department of Physics, UIUC, Urbana, IL61801,
(1998).  This is a Latex version of the original text.
}
\\ \small Department of Mathematics
\\ \small Southern Illinois University, Carbondale, IL62901
\\ \small jkocik{@}siu.edu  }
\date{}

\maketitle

\begin{abstract}
\noindent
\small 
The Ehresmann connection on a fiber bundle that is not compatible with a
(possible) Lie group structure is illustrated by the geometry of a general
anholonomic observer in the Minkowski space.  The 3D split of
Maxwell's equations induces geometric terms that are the (generalized)
curvature and torque of the connection.
The notion of torque is introduced here as a Lie coalgebra-valued
endomorphism field and measures the deviation of a connection from being
principal.
\\[3pt]
\\[5pt]
\end{abstract}



Notation:
$\XX M$ and $\La M$ denote
the $\FF M$-module of smooth vector fields and of exterior differential
forms on manifold $M$, respectively.
$\DD^kM$ denotes the set of smooth geometric $k$-distributions, i.e.,
fields of $k$-planes on $M$.


\section{Motivation}

The standard description of the {\it connection} in a fiber bundle
relies typically on some {\it group structure} acting along the fibers.
In particular, the connection form assumes values in the corresponding
Lie algebra, as does the curvature bi-form.
This group-based understanding of connection obscures its geometric
meaning and excludes possible applications of this concept in cases
where no particular symmetry is distinguished or demanded.

Although a general Ehresmann connection on a fiber bundle on
which {\it no} Lie group structure is defined or considered has already been
formulated (see e.g., Refs. 1 and 2).
Without a group structure, the connection form is described in terms
of an endomorphism field, and the curvature
tensor can be reinterpreted in terms of Fr\"olicher-Nijenhuis bracket.
Section 3I reviews these concepts in a format applicable to physics.
We also introduce a notion of a {\it torque} of connection that
measures the deviation of the connection from being principal.

The concept of a general connection seems interesting and should find
application in field theory and other areas of mathematical physics.
We illustrate the general connection with the anholonomic observer in
the Minkowski space.
Also, we describe the induced split of Maxwell's equations in a
coordinate-free manner (cf. Ref. 3),
which is another motivation for these notes
We show that the language of generalized connection allows
to preserve the elegance of Maxwell's equations in 3D
for anholonomic reference frame via the concepts of
``curvature" and ``torque" of the connection determined by an observer.

\section{Observer in the Lorentzian manifold}

Let $M$ be a Lorentzian space-time, i.e. a 4-dimensional pseudo-Riemannian
(possibly curved) manifold with metric tensor $g$ of signature (+ -- -- --).
An {\it observer} is a congruence of time-like curves in $M$
(called sometimes a ``perfect fluid"${}^7$. It should not be confused
with an {\it individual} observer ---  represented by a single time-like
curve in $M$).
An Observer can be equivalently represented by a normalized future-oriented
vector field $T\in\XX M$, $|T|^2=g(T,T)=1$.
(See e.g., Ref. 4).
The field of directions $\TT=\hbox{span}\,\{T\}$ will be called {\it time distribution} on $M$.
Define a one-form
$$
   \tau=g(T,\;\cdot\;) \quad \in\La^1M
\eqno(2.1)
$$
Distribution $\SS=:\Ker\tau$ will be called a {\it spatial} distribution
of observer $T$. Thus a choice of an observer amounts to fixing a pair
of two transversal distributions $\TT\in\DD^1M$ and $\SS\in\DD^3M $
that are mutually orthogonal and span the tangent space at each point of $M$
$$
        \TT\ \bot\ \SS \qquad\hbox{and}\qquad \TT\oplus\SS = TM
\eqno(2.2)
$$
or, using $g$, to fixing a pair $\{T,\tau\}$, such that
$$
\begin{aligned}
    \hbox{\caps time}  &= \TT = \hbox{span}\, T    \\
    \hbox{\caps space} &= \SS = \Ker\tau  \,.
\end{aligned}
$$
If the distribution of local space-hyperplanes $\SS$ is integrable, the
observer is called {\it holonomic}; otherwise it is
{\it anholonomic}.
Note that, by the Frobenius theorem, the space distribution
$\SS$ is integrable if $\tau\w d\tau=0$
(which in coordinates becomes a rather index-trashed equation,
$g_{ab} T^a \q_i (g_{jk}T^j) +\hbox{cycl}\;(b,i,k) = 0)$.
In this paper, we shall deal with the most general observer; in
particular, we shall see how anholonomic observers perceive the world of
electrodynamics.
\\

The pair $(T,\tau)$ can be used to construct on space-time an
endomorphism field, i.e. (1,1)-variant tensor field
$$
       \varkappa = \tau\otimes T .
\eqno(2.4)
$$

\begin{proposition}
The observer endomorphism field $\varkappa$ satisfies
$$
\begin{aligned}
    (i)\quad&  [T,\varkappa] = (\Ld_T\tau) \otimes T \\
    (ii)\quad&  [\varkappa,\varkappa] = -2 i_T(\tau\w d\tau) \otimes T
\end{aligned}
$$
where the brackets represent Fr\"olicher-Nijenhuis products of
vector-valued forms.
\end{proposition}

~\\
{\caps Proof:}
The definition of the Fr\"olicher-Nijenhuis bracket for any pair of
vector-valued forms is given in Appendix A.
Here we need special cases: for a vector field $T$, it is
$[T,\varkappa]=\Ld_T\varkappa$, and (i) follows directly via the Leibniz rule.
In the case of two
endomorphism fields (vector-valued 1-forms) $K$ and $L$, the
Fr\"olicher-Nijenhuis bracket $[K,L]$ is a vector-valued biform such
that for any two vector fields, $X$ and $Y$, it is (Refs. 5,6)
$$
[K,K] (X,Y) = 2[KX,KY] - 2K[KX,Y] - 2K[X,KY] + 2KK[X,Y] .
\eqno(2.6)
$$
Substituting (2.4) leads to (ii)  (use Eq. (A.11) of Appendix).
\QED
\\

Now, we shall interpret a Lorentzian space-time with an observer as a fiber
bundle with a connection.
Let $\sim$ denote the equivalence relation of belonging to the same
integral curve of $\TT$.  Define the {\it seeming space} of observer $T$
as the three-dimensional manifold of equivalence classes of $\sim$,
i.e., $S=M/\sim$. View space-time $M$ as a fiber bundle over $S$ with a
natural projection denoted
$$
          \pi: M\ \zzz\ S
\eqno(2.7)
$$
and equipped with a {\it connection} given by the `horizontal'
distribution $\SS$.
The connection form coincides with $\varkappa = T\otimes \tau$ defined above.
Clearly, $\varkappa(\SS)=0$ and $\varkappa(T)=T$.

\begin{figure}[H]
\centering
\includegraphics[scale=.7]{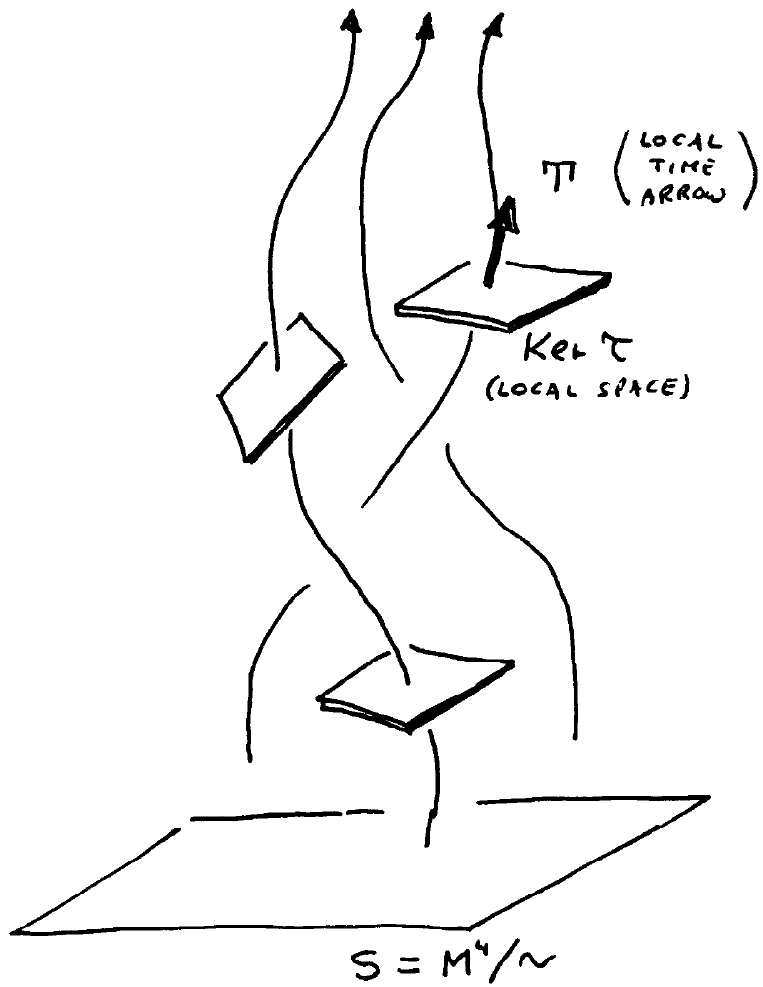}
\caption{An observer in space-time}
\label{fig:makaron}
\end{figure}

~\\
{\bf Remark:}
The fiber bundle (2.7) actually {\it is} a principal fiber bundle:
the group $\{\exp tT\}$ defines action via transport along the integral
curves of $T$.
However, this group action is not---in general---congruent with the
horizontal distribution; that is, $\varkappa$ is not a principal connection.
The magnitude of the deviation from being principal can be measured by
the Lie derivative of $\varkappa$ along $T$; call it the {\it torque} of the
observer (or the connection $\varkappa$):
$$
         \Torq(\varkappa)  = (\Ld_T\tau) \otimes T 
\eqno(2.8)
$$
Any space-like 3-dimensional manifold in $M$ may be viewed as
a section $\Psi:S\to M$.
Then the covariant derivative $\nabla\Psi$ measures `incompatibility' of
$\Psi$ as a candidate for a `space' with respect to observer $T$.
In particular, notice that $\nabla_v\Psi=g(T,\Psi_*v)$.
The next section will allow us to define the torque and the curvature of a
connection $\varkappa$ via the Nijenhuis bracket, so that
$$
\begin{aligned}
         \Torq(\varkappa)  &= [T,\varkappa]    = (\Ld_T\tau) \otimes T \\
         \Curv(\varkappa)  &= {\scriptstyle{1\over 2}}[\varkappa,\varkappa]
                                         = -i_T(\tau\w d\tau) \otimes T
\end{aligned}
\eqno(2.9)
$$
These  should not be confused with the standard curvature defined for
the covariant derivative in the Lorentzian manifold.

%



\section{Ehresmann Connection on a general fiber bundle}

This section contains a brief exposition of the concept of a general
Ehresmann con\-nec\-tion${}^1$
on a fiber bundle in a group-free context is presented.
A concept of connection torque is proposed to account for a
connection on a principal fiber bundle that is not group-invariant.

~\\ 
{\bf A. Connection form.}
Consider a fiber bundle $\{\pi: E\to M\}$ over a $n$-dimensional manifold $M$
with $\dim E=n+N$.
A {\it connection} on a fiber bundle $\{\pi:\; E\to M\}$ is any
$n$-distribution $H\subset\DD^n\!E$
$$
               H: \  p\ \zzz\  H_p<T_pE
\eqno(3.1)
$$
of subspaces complementary to the fibers.
Consequently, the connection determines a decomposition of tangent spaces
$$
                T_pE \cong H_p\oplus V_p
\eqno(3.2)
$$
at each $p\in E$, where $V\in \DD^NE$, is the {\it vertical
distribution} defined $V_p=\{v\in T_pE\;|\;\pi_*v=0\}$.
In particular, each tangent vector $X\in TE$
may be uniquely decomposed into a `vertical' and a `horizontal' part:
$
                         X= X_H + X_V
$.
Once chosen, $H$ (not necessarily integrable) is be called the {\it
horizontal distribution}.
\\

Splitting (3.2) can be described in terms of an endomorphism on $E$,
namely the {\it projection} of tangent vectors of $T_pE$
onto subspace $V_p$ along subspace $H_p$.  This is a linear map 
$\varkappa:T_pE\to T_pE$ such that
$$
\begin{aligned}
    (i) &\qquad \varkappa \circ \varkappa = \varkappa \\
   (ii) &\qquad \varkappa\Big|_H=0 \quad\hbox{and}\quad\varkappa\Big|_V=\rmid
\end{aligned}
\eqno(3.3)
$$
In particular, $\varkappa(X)=X_V$. Operator $\varkappa$ has been called somewhat improperly
a `connection form,' although it is  a  $(1,1)$-type tensor field and
will hence be termed a `field of endomorphisms.'

\newpage
\noindent
{\bf B. Lifting vectors.}
The horizontal distribution defines the {\it lifting} of vectors
from the basis manifold $M$  to the total space $E$;
for any point $p\in E$ and $m=\pi(p)\in M$, define a linear map
$\widetilde{\ \ } :\ T_mM \to T_pE$ such that
$$
\begin{aligned}
          (i)  &\quad \widetilde v\in H \\
          (ii) &\quad \pi_*(\widetilde v)=v
\end{aligned}
\eqno(3.4)
$$
for any $v\in TM$.
This can be extended to vector fields on $M$, which are lifted
to (horizontal) vector fields on $E$.
The definitions may be summarized in terms of the following
{\it exact} sequence
$$
0\ \zzz\ T_{\pi(p)}M\ \zzu{\widetilde{\ }}\ T_pE\ \zzu{\varkappa}\ T_pE
\ \zzu{\pi_*}\ T_{\pi(p)}M\ \zzz\ 0 .
\eqno(3.5)
$$
The extensions of the two operations (i) and (ii) to the vector fields
will be denoted by the same symbols:
$$
\begin{array}{ccc}
(i) &\qquad\hbox{lifting}\hfill   &\widetilde{\ \ }:\ \XX M \ \zzz\ \XX_HE  \\
(ii)&\qquad\hbox{projection}\hfill&\ \varkappa       :\ \XX E\ \zzz\ \XX_VE
\end{array}
$$
where $\XX_VE$ and $\XX_VE$ denote {\it vertical} and {\it horizontal}
vector fields, respectively.

~\\ 
{\bf C. Covariant derivative of a section.}
A horizontal distribution defines uniquely a {\it covariant derivative}
of a section $\Psi:M\to E$ along vector $v\in T_mM$ as a vertical vector
$$
        \nabla_v\Psi = \varkappa(\Psi^*(v))    \in V_{\Psi(m)}
\eqno(3.6)
$$
at point $\Psi(m)\in \EE$. (Equivalently,
$
             \nabla_v\Psi = \Psi_*v - \widetilde v
$,
where $\widetilde v$ is understood as lifted to the points of section $\Psi(M)$.)
In general, a {\it covariant derivative} of section $\Psi$ may be viewed
as a linear map $T_mM\to T_{\Psi(m)}E$ defined
$$
      \nabla\Psi \ = \ \varkappa\circ\Psi_*  .
$$

\begin{figure}[H]
\centering
\includegraphics[scale=1.2]{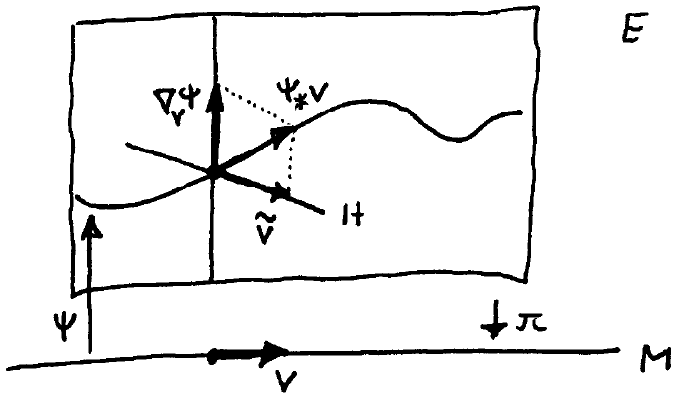}
\caption{Connection and covariant derivative}
\end{figure}

~\\
This, clearly, extends to vector fields and gives a (point-wise linear) map
$$
            \nabla\Psi:\ \XX M\to \XX E\Big|_{\Psi(M)}
\eqno(3.7)
$$
from vector fields on $M$ into vertical vector fields  along
the image $\Psi(M)$ of the section.

~\\
{\bf D.  Curvature via Nijenhuis bracket.}
Since the endomorphism $\varkappa$ may be viewed as a vector-valued
differential 1-form, the Fr\"olicher-Nijenhuis theory of differential
`vector forms' (i.e., $(1,k)$-type tensors) applies to it.
In particular, curvature may be defined in a coordinate-free, geometric
fashion.${}^2$

\begin{definition}
If $\varkappa$ is the (1,1)-tensor of connection, then
the curvature is a (1,2)-tensor defined as
$$
         \Omega ={1\over 2} [\varkappa,\varkappa] .
\eqno(3.8)
$$
where $[\;,\;]$ denotes the Nijenhuis bracket of vector-forms (see Appendix A).
\\

It is easy to see that the form $\Omega$ is vertical,
$i_v\Omega=0$ for any vertical $v$.  One can thus define a bi-form
$\omega$ on $M$ that assumes values in the vertical vector fields on $E$,
$$
              \omega(v,w) = \Omega(\widetilde v, \widetilde w)  .
\eqno(3.9)
$$
If $X,Y\in \XX M$ are two global vector fields on $M$, then
$\omega(X,Y)$ is a global (vertical) vector field on $E$.
\end{definition}

Since vector-forms form a graded Lie algebra, the Bianchi identity follows
immediately${}^2$
$$
            [\Omega, \varkappa] = 0  .
\eqno(3.10)
$$
Indeed, $[\Omega,\varkappa] = [[\varkappa,\varkappa],\varkappa] = 0$ due to the
(graded) Jacobi identity.
This pleasant result shows that the lack of a particular group
structure in the fibers is not an obstruction to the existence of quite a lot
of structure in the notion of a general connection.

~\\ 
{\bf E. Principal fiber bundle and torque.}
Let $\pi:E\to M$ be a principal fiber bundle, i.e., let a group $G$ act
freely and transitively on fibers of $E$.
In particular, the Lie algebra $L$ of $G$ is represented by the vertical
vector fields
$$
           \rho:\ L\zzz \XX_vE .
\eqno(3.11)
$$
induced by the diffeomorphisms of the action of the group.
The connection $\varkappa$ is
{\it principal} if it is invariant under the action of the group.
Consider a general Ehresmann connection on a principal
fiber bundle that does not necessarily agree with the group action.

\begin{definition}
{\it Torque} of a general connection on a principal fiber bundle is a tensor
$$
               \Torq(\varkappa) \in L^*\otimes T^{(1,1)}
\eqno(3.12)
$$
which, evaluated on $a\in L$, is
$$
               \Torq(\varkappa) (a) = \Ld_{\rho(a)}\varkappa .
\eqno(3.13)
$$
\end{definition}

Torque is clearly linear in $a$, and measures the deviation of the
connection from being principal.
If $\{e_p\}$ is a basis in $L$, and $\{\varepsilon_p\}$ in the dual space $L^*$,
then
$$
\Torq(\varkappa) = \varepsilon_p\otimes \Ld_{\rho(e_p)} \varkappa
\eqno(3.14)
$$
(summation over $p$).

\section{Geometry of Maxwell equations}

This section analyzes the geometry of Maxwell's equations
in the context of a (generalized) observer.

~\\
{\bf A. Maxwell Equations in $M^3$ and $M^4$}.
\\

Let us review the geometric content of Maxwell's equations as presented
in the standard way (see also Refs. 7, 8).

In the ``pre-relativistic'' formulation of Maxwell's equations, space is
a 3-dimensional manifold $M^3$ equipped with a Riemannian structure $g$,
while time appears in the theory as a {\it parameter} rather than a coordinate.
$$
\begin{aligned}
dE^\star  &= \rrho\\
             dE    &= -\q_t B  
\end{aligned}
  \qquad\qquad
\begin{aligned}
    dB   &=  0      \\
    dB^\star &= J + \q_t E^\star 
\end{aligned}
\eqno(4.1)
$$
where the fields $E$ and $B$ are differential 1-form and 2-form
respectively.  The Hodge star $\star$ acts as a linear map
$\star:\Lambda^kM\to\Lambda^{3-k}M$ and is defined by
$$
           \star\;(\alpha,\beta) = \alpha\w \star\beta
\eqno(4.2)
$$
where the bracket denotes the scalar product of forms (of the same
degree) induced from $g$ in $M^3$.
An immediate implication of Maxwell's equations is the `continuity
equation,'  $dJ + \q_t\rrho = 0$.
\\

In the relativistic formulation,  the Maxwell equations are rewritten
in 4-dimensional space-time, a manifold $M^4$ equipped with a
pseudo-Riemannian  structure $g$ of signature (+~--~--~--).
An {\it electromagnetic field}  is a differential bi-form
$F\in\La^2M^4$, and the {\it current-charge density} is a differential
3-form $j\in\La^3M$. The Maxwell equations are:
$$
\begin{aligned}
  dF  &= 0 \\
 d*F  &= j 
\end{aligned}
\eqno(4.3)
$$
The first equation implies via the Poincar\'e Lemma the (local)
existence of a differential 1-form $A\in\La^1M^4$, such that
$F=dA$. The second equation implies that \ $dj=0$ (continuity equation).
The laws of electromagnetism may be summarized in the following
diagram:
$$
\begin{array}{ccccccc}
           a & \zzzu d & F & \zzzu d & 0  \cr
            &~\cr
             &         &\dddr *           \cr
            & ~\cr
             &         &*F & \zzzu d & j & \zzzu d 0 
\end{array}
\eqno(4.4)
$$
where
$$
\begin{aligned}
    F  \in \La^2M  & \qquad \ot \ \hbox{electromagnetic field    }  \cr
G= *F  \in \La^2M  & \qquad \ot \ \hbox{dual electromagnetic field}  \cr
    A  \in \La^1M  & \qquad \ot \ \hbox{electromagnetic potential} \cr
    j  \in \La^3M  & \qquad \ot \ \hbox{charge-current }
\end{aligned}
$$
The Hodge star $*$ is now defined as a point-wise linear map
$\Lambda^kM\to\Lambda^{4-k}M$, such that
$$
           *\;(\alpha,\beta) = \alpha\w *\beta
\eqno(4.5)
$$
for any two forms of the same degree, where $(\;,\;)$ denotes he
pseudo-Euclidean scalar product on $M^4$.
\\
\\
{\bf B. An observer and fields}
\\
\\
Let $A$ and $\alpha$ be a vector field and a differential 1-form,
respectively,  $\alpha\in\La^1M$ and $A\in\XX^1M$. Then for any k-form
$\omega\in\La^kM$, the following formula holds (see e.g., Ref. 9):
$$
          A\jjj (\alpha\w\omega) + \alpha\w (A\jjj \omega )
          = \alpha(A) \cdot \omega .
\eqno(4.6)
$$
Denoting the {\it interior} and {\it exterior}
multiplications by
$$
\begin{aligned}
          i_A\omega        =& A\jjj\omega=\omega(A,\ldots) \\
          e_{\alpha}\omega =& \alpha\w\omega  
\end{aligned}
$$
we may rewrite (4.6) as $
       i_X\circ e_{\alpha} + e_{\alpha}\circ i_A
       \ = \
       \lambda\alpha,X\ra\cdot\rmid$.
If form $\alpha$ and vector $A$ are chosen so that $\alpha(A)=1$, then
one has a `decomposition formula'
$$
       i_X\circ e_{\alpha} + e_{\alpha}\circ i_A \ = \ \rmid
\eqno(4.7)
$$
that allows an exterior form to be split into two
parts---one that is ``parallel" to the direction of $A$ and one that is
``parallel" to the complementary direction of the co-plane spanned by the
kernel of $\alpha$, $\Ker\alpha$.
\\

Now,  consider a Lorentzian space-time with an observer $\{T,\tau\}$,
with $\tau(T) = 1$.  The split formula
$$
         e_{\tau}\circ i_T  + i_T\circ e_{\tau}  = \rmid
$$
allows any exterior form $\omega$ in space-time to be split
into a sum of two terms characteristic for a particular observer, i.e.
$$
\begin{aligned}
     \omega =&  e_{\tau}\circ i_T \omega + i_T (\tau\w \omega)r
            =  \omega_{time} + \omega_{space}  \cr
            =&  \tau\w \omega_T + \omega_S      
\end{aligned}
\eqno(4.8)
$$
where both $\omega_T$ and $\omega_S$ are purely spatial, since they
vanish under $i_T$.
Applying the split to the bi-form $F$ of an electromagnetic field gives
the observers' electric and magnetic component:
$$
           F = F_e + F_m \qquad\qquad\quad \hbox{where}\qquad
     \begin{cases}
              F_e = e_{\tau}\circ i_T F\equiv \tau\w(T\jjj F)\cr
             F_m = i_T\circ e_{\tau} F\equiv T\jjj(\tau\w F) . 
\end{cases}
$$
The observer perceives the magnetic field defined as a bi-form
$B=i_T(\tau\w F)$ and the electric field defined as 1-form $E=-i_T F$,
and the original tensor can be reconstructed as $F=E\w\tau + B$.
A similar split occurs for other fields, as summarized below:

\newpage

~\\
\hrule

$$
\begin{array}{cccccc}
F = E\w \tau + B \hfill &\qquad\qquad &\hbox{where}\qquad
                        &E = -i_T F \hfill&\qquad
                        &B=i_T\tau\w F \\
G = \tau\w H + D \hfill &\qquad\qquad &\hbox{where}\qquad
                        &H = i_T G  \hfill&\qquad
                        &D=i_T\tau\w G \\
j = \rrho-\tau\w J \hfill &\qquad\qquad &\hbox{where}\qquad
                        &J = -i_T j \hfill&\qquad
                        &\rrho=i_T\tau\w j \\
a = \varphi \tau + A \hfill &\qquad\qquad&\hbox{where}\qquad
                        &\varphi = i_T a \hfill&\qquad
                        &A=i_T\tau\w a 
\end{array}
$$

\hrule

~\\
~\\ 
{\bf C. An observer and the Maxwell equations}
\\
\\
Now we shall see the Maxwell equations under the
geometric split (4.8).  First, however, a definition of the two operators;

\begin{definition}
The spatial exterior derivative $d_3$ and the time derivative of a
differential form are, for observer $\{T,\tau\}$, the following
operators:
$$
\begin{aligned}
     \dot \omega &= \Ld_T\omega            \\
     d_3\omega   &= i_T\, \tau\w d\omega .
\end{aligned}
\eqno(4.9)
$$                    
\end{definition}

\begin{lemma}
In the reference system of observer $\{T,\tau\}$, a differential
equation $d\omega=\sigma$ involving exterior forms splits into a pair of
equations
$$
\begin{aligned}
                  d_3\omega_S   &=  \sigma_S + \Curv(\tau)\jjj \omega \\
  -\dot\omega_S + d_3\omega_T   &=  \sigma_T - \Torq(\tau)\jjj \omega 
\end{aligned}
\eqno(4.10)
$$
where
$$
\begin{aligned}
    \Torq(\tau)\jjj \omega =& (\Ld_T\tau)\w i_T \omega
                           =  \dot\tau\w\omega_T\cr
    \Curv(\tau)\jjj \omega =& -i_T(\tau\w d\tau)\w i_T \omega
                           =  -i_T(\tau\w d\tau)\w \omega_T . 
\end{aligned}
\eqno(4.11)
$$
The first equation of (4.10) is purely spatial (vanishes under $i_T$), and
the other is temporal (vanishes under $e_\tau$).  
\end{lemma}

\noindent
{\bf Proof:} Consider $d(\tau\w i_T \omega + i_T\tau\w \omega) = \sigma$.
Apply $i_T\w\tau$ to both sides to obtain the first equation:
$$
\begin{aligned}
 \sigma_S &= i_T (\tau\w d (\tau\w i_T \omega+i_T\tau\w \omega))\cr
          &= i_T (\tau\w (d\tau\w \omega_T-\tau\w d\omega_T+d\omega_S)\cr
          &= i_T (\tau\w d\tau)\w \omega_T+ i_T(\tau\w d\omega_S)\cr
          &= -\Curv\jjj \omega_T+ d_3\omega_S
\end{aligned}
$$
To obtain the second equation, apply $i_T$ to both sides:
$$
\begin{aligned}
 \sigma_T &= i_T d (\tau\w i_T \omega+i_T\tau\w \omega)\cr
          &= i_T (d\tau\w \omega_T -\tau\w d \omega_T +d\omega_S)\cr
          &= (i_T d\tau)\w \omega_T -i_T(\tau\w d\omega_T) +i_Td\omega_S)\cr
          &= (\Ld_T\tau)\w \omega_T -d_3\omega_T) +\dot\omega_S)
\end{aligned}
$$
where we used $\Ld_T=i_Td+di_T$.
\QED

\begin{theorem}
Maxwell's equations in the reference system of observer $\{T,\tau\}$ are
$$
\begin{aligned}
             d_3E   &=  -\dot B - \Torq(\tau)\jjj F \cr
              d_3B  &=   \Curv(\tau)\jjj F  
\end{aligned}
\qquad
\begin{aligned}
             d_3H  =& \dot D + J + \Torq(\tau)\jjj G \cr
             d_3 D =& \rho + \Curv(\tau)\jjj G  
\end{aligned}
\eqno(4.12)
$$
where
$$
\begin{aligned}
         \Torq(\tau)\jjj \omega =& (\Ld_T\tau)\w i_T \omega \cr
         \Curv(\tau)\jjj \omega =& -i_T(\tau\w d\tau)\w i_T \omega.
\end{aligned}
\eqno(4.13)
$$ 
\end{theorem}

~\\
{\bf Proof:} Apply Lemma 1 to both equations of (4.3).
\QED

~\\

From (4.12) follows that $\Curv(\tau)\jjj F$ can be interpreted as an
``apparent" magnetic charge, and $\Torq(\tau)\jjj F$ as an ``apparent"
magnetic current.  Similarly, $\Curv(\tau)\jjj G$ and
$\Torq(\tau)\jjj G$ contribute to the effective electric charge and
current, respectively. Clearly, if $d\tau=0$, then (4.12) reduces to the
standard 3D Maxwell equations.

The continuity equation and the potential can be treated similarly.

\begin{corollary}
The continuity equation for a general observer
becomes
$$
                  \dot \rrho + d_3 J = -\Torq(\tau) \jjj j
\eqno(4.14)
$$
\end{corollary}

~\\
{\bf Proof:} Write $j= \tau\w (-J)+\rho$ and apply Lemma~1.
There is only one 3D equation, since $dj$ is a 4-form and $\tau\w$ kills it.
\QED

\begin{corollary}
The equations for a potential are, for a general observer
$$
\begin{aligned}
               E &= - \dot A + d_3 \varphi - \Torq \jjj a  \cr
               B &= d_3  A - \Curv \jjj a        .    
\end{aligned}
\eqno(4.15)
$$
\end{corollary}

~\\
{\bf Proof:} Write $F=\tau\w(-E)+B$ and apply Lemma~1.
\QED

\begin{definition}
The reduced (spatial) Hodge star for an observer $(T,\tau)$ is defined as
$$
                          *_3 = i_T \circ *
\eqno(4.16)
$$
where $*$ denotes the Hodge star in space-time.  
\end{definition}

\noindent
{\bf Remark:}  If $d\tau=0$ and $\imath:S\subset M$ is a 3-dimensional
submanifold such that $\imath^*\tau=\hbox{const}$  (``instantaneous space"), then
the reduced spatial Hodge star (6.16) coincides with the Hodge star
$\star$ on $S$, determined by the induced metric $\imath^*g$, that is,
$\star\circ\imath^* = \imath^*\circ *_3$.

\begin{proposition}
The Hodge star $*$ intertwines the operators of $e_\tau$ and $i_T$:
$$
\begin{aligned}
   i_T \circ * &= (-)^k * \circ e_\tau  \cr
         e_\tau \circ * &= (-)^{k+1} * \circ i_T 
\end{aligned}
\eqno(4.17)
$$
where $k$ is the degree of the form on which the product acts, and we
abbreviate $(-)^k=(-1)^k$.   
\end{proposition}

~\\
{\bf Proof:}  Let us show the first relation:
$$
\begin{aligned}
* \circ e_\tau \omega &= * (\tau\w\omega) = (g^{-1}(\tau\w\omega))\jjj \eta  \cr
                    &= (T\w\Omega)\jjj \eta = (-1)^k (\Omega\w T)\jjj \eta \cr
                    &= (-)^k i_T\circ i_\Omega \eta  =  (-)^k i_T * \omega 
\end{aligned}
$$
where $\Omega = g^{-1}(\omega)$ and $k=\deg\omega$. The second relation
follows similarly.
\QED

\begin{corollary}
For any observer $(T,\tau)$, the Hodge relation $*F=G$ reduces to
$$
           *_3E = D  \qquad\hbox{and}\qquad *_3H = B
\eqno(4.18)
$$
\end{corollary}

~\\
{\bf Proof}:  Indeed, using (4.14), one calculates
$*_3E = i_T*E = -i_T*i_TF = i_T e_\tau * F = i_Te_\tau G = D$.
Similarly for a magnetic field,
$*_3B = i_T*B = i_T*i_Te_\tau F = i_T e_\tau * e_\tau F = i_Te_\tau i_T *
F = i_T * F = i_T G =  H$.
\QED

Note that the operators $i_T$ and $e_\tau = \tau\w$ \  satisfy the
interesting algebraic property
$$
      i_T^{} \circ e_\tau \circ i_T^{}  = i_T^{}
      \qquad\hbox{and}\qquad
      e_\tau \circ i_T  \circ e_\tau  =  e_\tau
\eqno(4.19)
$$
(cf., Temperley-Lieb algebra or Artin braid group).
\\


\section*{Appendix:   Nijenhuis bracket}

Consider endomorphisms of the space of exterior differential forms.
A graded map $a:\Lambda M\to\Lambda M$ is of degree $\deg a \in\mathbb Z$, if
$a:\Lambda^k M\to \Lambda^{k+\deg a}M$.
Define the following product of graded maps in $\La M$
(see e.g., Refs. 10, 11, 12, 6):
$$
     \sprl a, b\sprr = a\circ b-(-)^{(\deg a\cdot\deg b)}\ b\circ a
\eqno(A.1)
$$
and extend it by linearity to the space spanned by graded maps of
$\Lambda M$.  These immediate properties
$$
\begin{aligned}
     (i)\quad& 
                \hbox{(bi-linearity)} \cr
     (ii)\quad& \sprl a, b\sprr = -(-)^{(\deg a\cdot\deg b)}\ \sprl b,a\sprr
               \qquad \hbox{(super skew-symmetry)}  \cr
     (iii)\quad& (-)^{(\deg a\cdot\deg c)}\ \sprl a,\sprl b,c\sprr\sprr
             + \hbox{cyclic terms} =0
               \qquad \hbox{(Jacobi identity)} 
\end{aligned}
\eqno(A.2)
$$
turn the space of graded maps into a Lie superalgebra  
with superbracket $\sprl\;,\;\sprr$.
Endomorphism $D$ is called a {\it derivation}, if it satisfies the
graded Leibniz rule
$$
D(\alpha\w\beta) = D\alpha\w\beta
                 + (-)^{\deg D\cdot\deg\alpha}\alpha\w D\beta
\eqno(A.3)
$$
The well-known examples include the exterior derivative and the
contraction with a vector field
$$
        \La^kM\zzu{d}\La^{k+1}M
                   \qquad \hbox{and}\qquad
        \La^kM\zzu{i_X}\La^{k-1}M
\eqno(A.4)
$$
where $X\in \XX M$.
Another derivation can be defined by contraction
with any vector-form (vector-valued form), i.e., an element of
$$
   \XX^1 M\otimes\La^k M \subset T^{1,k}M .
$$
Indeed, for a homogeneous vector-form
$$
        A=\alpha\otimes A   \qquad \alpha\in\La^kM\qquad A\in\XX M
$$
the contraction defined as
$$
i_A\omega = \alpha\w i_A\omega
\eqno(A.5)
$$
is is a derivation of degree $(k-1)$, which can be extended by
linearity to a general vector-form.

The set $Der\;\La M$ of all derivations is closed under the product of
the superbracket.   Define a Lie derivative along a vector-form $A$ as
$$
       \Ld_A = \sprl i_A, d \sprr .
\eqno(A.6)
$$
Since derivations form a superalgebra, $\Ld_A$ is a derivation.
By the same argument, the superbracket of two Lie derivatives is a
derivation.  It can be shown that for any
two vector-forms $A$ and $B$, there exists a vector-form denoted
$[A,B]$, such that
$$
       \sprl\Ld_A,\Ld_B\sprr =
                \Ld_{[A,B]}
\eqno(A.7)
$$
The vector-form $[A,B]$ is called a Nijenhuis (or Fr\"olicher-Nijenhuis)
bracket of $A$ and $B$.
The following theorem generalizes this result${}^6$:

\begin{theorem}
(Michor) Any derivation $D\in Der_k\La M$ is of form
$$
   D= \Ld_K+i_L
\eqno(A.8)
$$
for some vector-forms $K$ and $L$.  In particular, the exterior
derivative is $d=\Ld_{\rmid}$.  In addition
$$
\begin{aligned}
 (i)\quad& L=0 \qquad\hbox{iff}\qquad \sprl D,d\sprr=0 \cr
(ii)\quad& K=0 \qquad\hbox{iff}\qquad D|_{\FF M}=0 . 
\end{aligned}
$$
\end{theorem}

Since $\sprl \sprl \Ld_K,\Ld_L\sprr,d\sprr=0$ (by the Jacobi identity
(A.2.iii)), then, by the above theorem, there exists a vector-form $[K,L]$
such that
$$
          \sprl \Ld_K,\Ld_L\sprr = \Ld_{[K,L]} .
$$
For the special cases where $K$ and $L$ are two endomorphism fields
(vector-valued 1-forms), the Nijenhuis bracket $[K,L]$ is a
vector-valued biform, which, evaluated for two vector fields $X$ and $Y$, is
$$
\begin{aligned}
[K,L] (X,Y) = &\phantom{+} [KX,LY] - [KY,LX]\cr
              &-  L[KX,Y] + L[KY,X] \cr
              &-  K[LX,Y] + K[LY,X]  \cr
              &+ LK[X,Y]  + KL[X,Y] . 
\end{aligned}
\eqno(A.9)
$$
The Nijenhuis torsion of an endomorphism field $K$ is defined via the
Nijenhuis bracket of $K$ with itself:
$$
     N_K = {1\over 2}[K,K](X,Y) = [KX,KY] -K[KX,Y] -K[X,KY]+K^2[X,Y] .
\eqno(A.10)
$$
The Nijenhuis torsion of a homogeneous tensor $A\otimes\alpha$ is
$$
\begin{aligned}
   {\scriptstyle 1\over 2} [A\otimes \alpha ,\; A\otimes\alpha]
         & = A\otimes(\alpha\w\Ld_A\alpha - \alpha(A)\, d\alpha )      \cr
         & = A\otimes(\alpha\w d(\alpha(A)) - i_A\; \alpha\w d\alpha) . 
\end{aligned}
\eqno(A.11)
$$



\end{document}